\documentclass[pdflatex,sn-mathphys-num]{sn-jnl}

\usepackage{graphicx}%
\usepackage{multirow}%
\usepackage{amsmath,amssymb,amsfonts}%
\usepackage{amsthm}%
\usepackage{mathrsfs}%
\usepackage[title]{appendix}%
\usepackage{xcolor}%
\usepackage{textcomp}%
\usepackage{manyfoot}%
\usepackage{booktabs}%
\usepackage{algorithm}%
\usepackage{algorithmicx}%
\usepackage{algpseudocode}%
\usepackage{listings}%

\theoremstyle{thmstyleone}%
\usepackage{braket}

\usepackage{changepage}

\usepackage{tcolorbox}

\theoremstyle{thmstyletwo}%

\theoremstyle{thmstylethree}%

\begin{document}

\title[Article Title]{The Informational Observer in Relational Quantum Mechanics}


\author*[]{\fnm{Bethany} \sur{Terris}}\email{beth.terris@gmail.com}

\affil[]{\orgdiv{Université Paris-Saclay}, \orgname{CEA, LARSIM}, \orgaddress{\street{91191}, \city{Gif-sur-Yvette}, \country{France}}}



\abstract{Relational Quantum Mechanics (RQM) treats quantum states as observer-dependent facts rather than absolute properties. While this relational stance is conceptually attractive, it raises concerns about empirical confirmation, particularly in multi-observer scenarios. Existing responses within RQM focus on securing agreement between observers by strengthening the status, stability, or accessibility of recorded outcomes. However, they leave open a more basic question: what grounds the persistence of an observer across time? Scientific observation presupposes stable records and the capacity to relate outcomes across successive measurements. We argue that the minimal definition of the observer in RQM as a merely interacting physical system is insufficient to support this requirement. We propose a complementary account of the observer that distinguishes physical interaction from informational coherence, and show how this distinction supports empirical confirmation in Wigner's friend–type scenarios.}




\maketitle

\section{Introduction}

In the 1990s, Relational Quantum Mechanics (RQM) \cite{rovelli1994quantum,rovelli1996relational} was developed as a compelling alternative to more traditional interpretations of quantum theory such as the Copenhagen interpretation, the many-worlds interpretation \cite{everett1957relative,deutsch1985quantum}, or Bohmian mechanics \cite{bohm1952asuggested}. The core idea of RQM is that the quantum state does not provide an absolute description of a system. Instead, it provides a relational description: states are defined only relative to another system that functions as an observer.

The difference between the interpretations listed above and RQM can be illustrated by a simple example. At time $t_i$, a measurement is performed on a system in the spin basis, and at a later time $t_i+1$, an outcome is read out. In interpretations such as Bohmian mechanics or the Copenhagen interpretation, it is natural to say that the spin of the system is, for example, up at time $t_i$. Meanwhile, in Everettian quantum mechanics, the outcome is typically described as occurring within a particular branch of the universal wavefunction. RQM reformulates such claims in explicitly relational terms: the spin of the system is up relative to some observer $O$ that has interacted with the system at time $t_i$. The emphasis is thus placed on the relation between the system being measured and the system enacting the measurement.

RQM therefore explains quantum phenomena by rejecting absolutism and embracing an observer-dependent notion of states, or, in more recent formulations \cite{di2022relational}, facts. Within this framework, it is entirely consistent for different observers to hold different, yet equally valid, descriptions of the same system. An observer $A$ may describe the system as being in state $a$, while an observer $B$, in a different context, may describe it as being in state $b$. Both descriptions are valid relative to their respective relational contexts.

This observer-dependence does not render RQM a subjective interpretation \cite{rovelli2021relational}. Rather, RQM replaces the notion of an absolute state with a fundamental web of relations between physical systems, all of which are treated on equal epistemological footing. The observer is defined in RQM as any physical system that interacts with another system. RQM therefore dissolves the observer-system split by claiming no privileged class of observers.

However, this minimal definition of the observer raises significant philosophical concerns regarding what the process of observation involves and how purely physical systems can meaningfully function as observers. If observation is reduced to a single interaction, that is, to the establishment of a correlation between systems, then it is a momentary event, not a process.

Scientific practice, however, relies on more than isolated interaction events. It depends on an observer’s capacity to retain information and to form a coherent narrative about measurement outcomes across time. 
Without this, empirical confirmation would be impossible, as isolated outcomes could neither be compared nor assessed for consistency, undermining the repeatability and evidential basis of scientific inquiry.

Observation, in this sense, must therefore be understood as a temporally extended process rather than a momentary interaction. The system that registers an outcome must also be able to remember it and relate it to subsequent observations, thereby sustaining the continuity required for empirical assessment, a requirement already implicit in Everett's insistence that observers be understood as physical systems equipped with ``recording devices'' \cite[p.~9]{Everett1973}.

This raises a question: what ensures that the same observer persists across time? If an observer is nothing more than a physical system undergoing interactions, then its identity appears vulnerable to decoherence, physical disruption, or the gradual replacement of its constituent parts, analogous to the Ship of Theseus paradox \cite{Jansen2011-JANTSO-13}, and related to an expansive literature on problems of personal identity \cite{locke1847essay,parfit1971personal,parfit1976lewis,Parfit1984-PARRAP,Parfit1997-PARTUO,Parfit2012-PARWAN,Jenkins_2006}. In such cases, it is unclear what grounds the claim that the observer at one moment is identical to the observer at a later moment. Without this extension of identity, there is no stable system to which records are meaningful, and no way for the predictions of the interpretation to be compared and confirmed. This threatens empirical confirmation and, ultimately, the scientific viability of RQM.

Analogous concerns have been raised and addressed within the Everett interpretation, where proposed solutions include grounding observer persistence in decoherence-defined emergent branches \cite{wallace2012emergent}, positing determinate mental states that follow a single probabilistic trajectory through the branching structure \cite{albert1988interpreting}, and applying perdurantist accounts of identity in which observers are temporally extended entities rather than enduring substances \cite{saunders1995time}. However, these responses are tailored to the branching structure of the Everett interpretation and do not carry over straightforwardly to RQM.

Recent work within RQM has sought to address concerns about empirical confirmation by focusing on how agreement can be recovered across different observers. These approaches typically aim to explain how observer-dependent descriptions can nevertheless be reconciled once observers interact, for example by appealing to shared facts \cite{di2021stable,pienaar2021quintet}, internally consistent descriptions \cite{di2022relational}, or cross-perspective links (CPL) \cite{adlam2022information}. Such proposals are primarily motivated by cases involving a lack of intersubjective agreement, particularly in nested-observer scenarios such as Wigner’s friend–type experiments \cite{wigner1995remarks,frauchiger2018quantum,brukner2018no}.

However, these approaches leave unresolved a more basic issue concerning the observer itself. Even when cross-perspective links are specified, it remains unclear what grounds the persistence of an observer across time as a bearer of records. Indeed, as identified in the literature on CPL \cite{adlam2022does}, without stable observer identity the conditions required for record comparison and agreement cannot be satisfied. Existing approaches therefore expose a conceptual gap in RQM concerning observer persistence.

The main objective of this article is to propose a solution to this problem by rethinking what counts as an observer in RQM. We argue that the minimal physicalist definition is insufficient to support the processes required for observation through time. Instead, following a similar argument from Grinbaum \cite[Ch.~4.5]{grinbaum2004significance}, we propose a complementary definition of the observer, consisting of a physical role grounded in interaction, and an informational role grounded in coherence across time.

The article proceeds as follows. Section \ref{Sec:rethinking} addresses the minimal requirements for observation in scientific practice, distinguishing the physical observer from the informational one. This section emphasises that the observer should be fundamentally understood as time-delocalised. Section \ref{Sec:coherent} asks why coherence matters, developing sequential weak values as probes of informational coherence across time, from which the informational observer emerges.

Then, Section \ref{Sec:two-step} addresses the problem of intersubjective agreement in RQM. This section reviews the Wigner's friend set-up and explains how our redefinition of the observer can be supplemented to CPL to recover intersubjective agreement. This leads to increased confidence in the predictions made by the interpretation, providing a possible route towards empirical confirmation of RQM. This section also provides a limit case from an application of the proposal to the Frauchiger-Renner paradox. Finally, Section \ref{Sec:conc} summarises the argument, discussing its position within broader philosophical perspectives, and noting some future research directions.

\section{Rethinking the Observer}\label{Sec:rethinking}

\subsection{The problem of the minimalist observer in RQM}

The relational definition of the observer is intentionally minimalist and egalitarian; it avoids positing a privileged class of observers and dissolves the system-observer distinction that haunts other interpretations \cite{rovelli1996relational}. All physical systems, regardless of size or complexity, can serve as observers within this framework.

Here it is important to distinguish two different senses in which the notion of an observer appears in discussions of RQM. In the standard formulation of the interpretation, the observer plays a metaphysical role: an observer is simply the physical system relative to which facts are realised through interaction. In this sense, observers belong to the ontological structure of the theory itself, since different systems may stand in different networks of relational facts.

However, discussions of scientific observation and empirical confirmation often invoke a more epistemic notion of the observer. In this second sense, an observer is not merely a system relative to which facts occur, but a system capable of sustaining a coherent evidential perspective across time. These two notions are related, but they are not identical, and much of the present discussion concerns the second.

From this epistemic perspective, the minimalist account appears incomplete. Observation in scientific practice is not reducible to a single interaction event. It is an extended process involving the retention of memories, the comparison of outcomes, and the interpretation of information across time. A single physical interaction may establish a relational fact, yet still fail to support the kind of informational continuity required for empirical confirmation.

This claim aligns with well-established positions in the philosophy of science. Hacking emphasises that observation is inseparable from intervention and representation \cite{hacking1983representing}, requiring an extended engagement with phenomena. van Fraassen argues that observation is theory-dependent and gains significance only within a conceptual framework \cite[Ch.~1.3]{van1980scientific}, meaning that the outcomes of observation are only meaningful when structured by prior expectations and comparative assessment. Healey similarly argues that quantum states should be understood as tools for agents to coordinate their actions and update their beliefs rather than as descriptions of physical reality, implying that observation is meaningful only within an ongoing inferential practice that extends across time \cite{healey2017quantum}. Kuhn further shows that observations are interpreted through historically structured paradigms \cite[Ch.~X]{kuhn1997structure}, thus their meaning emerges only through comparison over time. From the perspective of confirmation theory, Reichenbach and Carnap both emphasise that observational outcomes function as evidence only insofar as they are retained, comparable, and integrated into a temporally extended inferential framework \cite{Richardson1938-RICEAP-4,carnap2014logical}. Together, these views support the claim that observation is not a momentary interaction, but a temporally extended process involving retention, comparison, and interpretation which is necessary for empirical validity.

Despite the existence of timeless formulations of RQM \cite{rovelli2011forget}, the intuition of observation as a momentary event is often smuggled into the theory. In this sense, RQM often focuses observation only on the event of interaction and thus the physicalist definition of the observer neglects the extended informational processes that turn a one-off event into a meaningful observation. This treatment of the observer means that there cannot be any retention or recall of prior events, and that there is no basis for confirmation of events or outcomes. The minimal account fails to explain how a system that engages in one interaction at an initial time $t_i$ and another at a final time $t_f$ can integrate the outcomes of the interactions into a coherent framework. Hence, it fails to account for what would typically be required of a system to function as an observer within the broader practices of scientific inquiry.

\subsection{Building a complementary account of the observer}

To count as an observer, we argue that a system must satisfy the following three \textbf{Conditions for Observerhood}:

\subsubsection*{Condition 1}

The system must be able to record outcomes: the information obtained in an interaction must be instantiated in some physical medium, whether in a memory register or as a macroscopic trace. Without recording, the event evaporates and cannot contribute to any further reasoning.

\subsubsection*{Condition 2}

The record must be retained. A fleeting imprint that vanishes immediately does not permit reasoning across time. This condition aligns with Di Biagio and Rovelli's proposal that facts must be stable \cite{di2021stable}. Without retention, there is no informational continuity on which to build comparison, inference, or empirical confirmation.

\subsubsection*{Condition 3}

The system must be able to relate records across time in a coherent manner. Even if individual outcomes are recorded and retained, they are not yet part of a single perspective unless they can be integrated into one diachronic informational thread. Without such integration, the system has only a set of isolated points, each epistemically disconnected from the rest. This coherence criterion is fully fleshed out in Section \ref{Sec:coherent}, where we argue that this ability to construct coherent narratives is what elevates a registering system to the status of an observer. Scientific practice relies on the comparison of outcomes across time and the evaluation of their fit. Without coherence, there is no continuous perspective and ultimately no basis for empirical inquiry.

\hfill

Condition 1 captures the `Observers Observe' postulate: that some physical systems register definite outcomes when they interact \cite{schmid2025copenhagenish}. However, 'Observers Observe' is silent on what happens to those outcomes after the moment of registration. Conditions 2 and 3 go beyond it, requiring that outcomes be retained and coherently integrated across time, which is precisely the structure that a purely interaction-based account of the observer leaves out.

So what do we need to add to the minimalist definition in order to fulfil the requirements for observation? The physical description is necessary as there must be a physical system interacting with another physical system in order to obtain facts or outcomes, fulfilling condition 1. However, we also need the observer to examine whether a coherent narrative has been retained by these records across time, fulfilling conditions 2 and 3.

The distinction introduced above between the metaphysical and epistemic roles of the observer can now be made more precise. The minimal interaction-based observer found in standard formulations of RQM primarily fulfils the metaphysical role: it is the physical system relative to which facts are established through interaction. However, the epistemic role required for scientific observation and empirical confirmation involves additional informational structure extending across time. Following Grinbaum \cite[Ch.~4.5]{grinbaum2004significance}, we articulate these two dimensions of observerhood in terms of two complementary functional roles: the P-observer and the I-observer.

\hfill

\begin{quote}
\textbf{P-observer:} This role is fulfilled by any system that physically interacts with another, giving rise to a fact. This fact is meaningful relative to the interacting system, reflecting the outcome of this specific encounter.
\end{quote}

\hfill

The P-observer resides within the quantum dynamics; it is momentary and interaction-based, defined entirely in terms of its physical interaction with another system. The P-observer is the role that is already recognised in the minimalist definition of the observer in RQM.

\hfill

\begin{quote}
\textbf{I-observer:} This role is fulfilled by a system that sustains and organises informational records across time, allowing outcomes of interactions to be retained, compared, and interpreted. The I-observer is not defined by a single physical interaction, but by the coherence of informational structure that persists through successive interactions.
\end{quote}

\hfill

The I-observer, or informational observer, was not a part of older formulations of RQM. However, more recently similar ideas have begun to be explored, such as the addition of postulates involving cross-perspective links, which require that all the information held by an observer be encoded in its physical variables \cite{adlam2022information}. Yet such proposals fall short of addressing how the informational structure is maintained within an observer across time.

The I-observer is introduced precisely to fill this gap. It captures what is missing from a purely interaction-based account: an enduring informational structure within which successive outcomes can be integrated. Rather than being associated with individual encounters, the I-observer is defined by the coherence of information across multiple interactions, making it possible to treat records as belonging to the same observer over time. In this sense, the I-observer underwrites the continuity required for scientific reasoning and empirical confirmation.

Such temporal coherence has previously been characterised in informational terms. For example, Grinbaum has argued that coherence across observations may be expressed via vanishing Shannon entropy \cite{grinbaum2010quantum}. In the present article, we approach the same structural requirement from a different perspective, focusing not on entropic measures but on how coherence across time can be assessed operationally within quantum theory.

The P-observer thus marks the occurrence of a fact by undergoing physical interactions with other systems, fulfilling condition 1 for observerhood. The I-observer makes that fact available to a coherent temporally extended perspective, fulfilling conditions 2 and 3. Thus, these two roles are equally necessary; without the P-observer, there could be no interaction, and without the I-observer, no fact would have enduring epistemic value.

Crucially, this is not a dualist model of the observer. The P- and I-observers are not to be interpreted as two separate entities, but as two functional roles that a single system must play in order to function as an observer. A simple qubit may act as a P-observer during an interaction, but it could only act as an I-observer and thus function as a full observer if the qubit is complemented by some additional degrees of freedom in which information could be retained.

The distinction between these roles reflects two complementary approaches to the question of what an observer is. The P-observer represents a constructive approach: it is defined by its physical constitution and the interactions it undergoes, so that the material substrate matters essentially. The I-observer represents an informational approach: what counts is not the physical hardware but the structural properties of the information it sustains across time, and the same informational role could in principle be realised by different physical substrates.

Grinbaum argues that this dual description of the observer is necessary due to the structure of information-theoretic reconstructions of quantum mechanics \cite{grinbaum2003elements,grinbaum2004significance,grinbaum2007reconstructing,grinbaum2007reconstruction}. These reconstructions start from a set of axioms about information which refer to an agent or observer who acquires, stores, and uses that information \cite{hardy2001quantum,clifton2003characterizing,chiribella2011informational}. Quantum theory itself does not tell us which part of the observing system counts as the `memory', or how the memory may be protected from disturbances or decoherence. This choice is a modelling decision made in the meta-theory, where we choose which subsystem's degrees of freedom will act as the memory record.

Interestingly, the I-observer is time-delocalised. Unlike the P-observer which is localised at a specific, discrete point in time (the moment of interaction), the I-observer is spread out across time. It is constituted in virtue of the informational coherence of the records. The I-observer is a structure that emerges when there is sufficient coherence, and it is the logical condition that allows measurement outcomes to be assigned to a single perspective.

In this sense, what counts as an observer for one interaction may not count as an observer for another interaction. If Alice's interaction with Bob does not form a coherent narrative, then Alice's I-observer role is non-existent, and she does not count as an observer. But if a coherent narrative is formed by another interaction between Alice and Charlie, then the I-observer emerges from this coherence and Alice counts as an observer in this context. Thus, the emergent nature of the I-observer leads to a relational definition of the observer.

It is important to clarify the scope of this proposal. The account developed here is not intended to resolve the stronger metaphysical question of whether a P-observer numerically persists across time in the sense required to underwrite a single continuous relational reality. That problem remains a substantial challenge for RQM. Rather, the present proposal concerns the informational conditions under which temporally separated records may function as belonging to a single coherent perspective sufficient for observation, memory, and empirical confirmation. In this sense, the emergence of the I-observer should be understood as a form of informational continuity grounded in coherent record structure, rather than as a proof of metaphysical identity across time.

\section{Structural Coherence from Sequential Weak Values}\label{Sec:coherent}

\subsection{Coherent records}\label{subsec:coherentrecords}

Given this redefinition of the observer, what we call an observer is more than a physical system that records facts. It is a system whose informational role persists through time. Observation thus becomes a pattern that arises when certain informational coherence conditions are met. But this coherence is not a trivial assumption; it must be demonstrable.

Having distinguished the P-observer from the I-observer, we can now specify the structural condition required for the emergence of the I-observer. We call this condition a \emph{coherent record}.

\begin{quote}
\textbf{Coherent record:}
A coherent record is a temporally extended sequence of stored outcomes that can be embedded within a single relational history relative to an observer's informational boundary conditions.
\end{quote}

The relevant boundary conditions are the initial informational state and the final informational state, otherwise respectively known as the pre- and post-selected states. These boundary conditions determine whether the recorded outcomes can be understood as belonging to a single history rather than a disconnected collection of events.

Formally, let
\[
H_R=\Pi_{r_n}(t_n)\cdots\Pi_{r_1}(t_1)
\]
denote the history operator associated with a sequence of outcomes $r_1,\ldots,r_n$. Let $\ket{\psi}_O$ and $\ket{\phi}_O$ denote the pre- and post-selected boundary states associated with observer $O$. We then define

\begin{equation}\label{eq:projformswv}
    \langle H_R \rangle^O_W \;=\; 
\frac{\langle \phi |\, H_R \,| \psi \rangle}{\langle \phi | \psi \rangle}.
\end{equation}
\hfill

Here, the superscript $O$ indicates that the weak value is evaluated relative to observer $O$. The boundary states $\ket{\psi}_O$ and $\ket{\phi}_O$ should therefore be understood as states of the observer's informational register (or memory degrees of freedom) within that observer-relative description. Likewise, the projectors composing $H_R$ represent recorded outcomes associated with that informational structure across successive times. Which physical degrees of freedom constitute the observer's informational register is itself a modelling choice made at the meta-theoretical level, as discussed in Section~\ref{Sec:rethinking}.

The quantity defined in equation~\eqref{eq:projformswv} evaluates the compatibility of a proposed record sequence with the observer's informational boundary conditions. This allows us to state our criterion for structural coherence:

\begin{tcolorbox}
\textbf{Coherence Criterion:}
A record $R$ is structurally coherent relative to observer $O$ iff
\[
\langle H_R\rangle_W^O \neq 0.
\]
Equivalently, the associated history operator admits a non-vanishing sequential weak value between the relevant boundary conditions.
\end{tcolorbox}
\hfill

Equation~\eqref{eq:projformswv} should therefore be understood as a structural consistency test. It evaluates whether a proposed sequence of recorded outcomes can be embedded within a single diachronic history compatible with the observer's informational boundary conditions. A non-vanishing value indicates that such an embedding is possible, whereas a vanishing value indicates that no continuous relational history exists connecting the recorded outcomes.

The formalism is therefore not intended to describe an observer-independent history of the measured system, but rather the structural coherence of records within an observer-relative informational framework. Accordingly, the use of a single observer-relative index $O$ throughout the coherence criterion should not be understood as presupposing the numerical identity of an observer across time, but instead as fixing the relational perspective within which that coherence is evaluated.

This notion of structural coherence is stronger than mere logical consistency. Logical consistency requires only the absence of explicit contradiction between propositions. Structural coherence, by contrast, requires that the recorded outcomes be jointly compatible with a single relational history relative to the observer.

To illustrate the distinction, suppose Alice records the following sequence: `spin-up at $t_1$', `spin-down at $t_2$', and `spin-up at $t_3$'. Taken as a list of propositions, this record is logically consistent: no statement negates another. However, if Alice’s boundary conditions do not admit any physical process (such as a spin-flip interaction) capable of connecting these outcomes between $t_1$ and $t_2$, then the history operator associated with this record yields a vanishing value in equation~\eqref{eq:projformswv} since the successive projectors cannot be jointly realised along a single evolving history. In this case, the record cannot be embedded into a single diachronic history and is therefore structurally incoherent.

On the other hand, if an appropriate dynamical channel is present, the same sequence of outcomes may yield a non-vanishing value since the intervening dynamics provides a relational link that connects the projectors at different times into a single admissible history compatible with the observer's boundary conditions. This non-vanishing value indicates that the record can be assigned to a coherently evolving structural narrative.


A macroscopic illustration makes the previous point more intuitive. Suppose instead that Bob brews a cup of tea at $t_1$, at $t_2$ the cup seems to contain coffee instead of tea, and at $t_3$ it contains tea again. Nothing \textit{logically} forbids this description since each claim is indexed to a different time. The propositions do not contradict one another, as neither makes an incompatible claim about the cup's contents at the same time. But unless Bob can situate the episode in a plausible history (for example, perhaps someone swapped the cup at $t_2$), the sequence cannot be embedded as one diachronic narrative in his perspective. It is thus structurally incoherent.

A coherent narrative is therefore more than a list of outcomes. It is a structured informational sequence that can be situated within one history. The observing system may say `I saw outcome $x$ at time $t_1$', and `I saw outcome $y$ at time $t_2$', but the existence of a coherent record allows her to also say: `Outcomes $x$ and $y$ belong to a single diachronic history compatible with my boundary conditions, and so they can be treated as parts of one perspective.'

This coherence criterion aligns with Rovelli's postulate that there is a maximum amount of `relevant information' obtainable from a quantum system \cite{rovelli1996relational}. Information is considered `relevant' when it contributes to an internally coherent informational structure, specifically one that can be embedded into a history rather than collapsing into disconnected fragments. This information is not merely epistemic, but structural \cite{grinbaum2003elements}; it is defined by the compatibility of an observer's descriptions over time. In this view, coherence ensures that classical records are structurally stable, the very conditions under which the I-observer emerges.

While the quantum formalism provides the mathematical tools for describing coherence, it does not by itself guarantee that an observer's records will be structurally coherent. Whether a given record can be embedded within a single history depends on the specific boundary conditions (the pre- and post-selected states), the choice of measurement, and the dynamical context that links them. In other words, the quantum formalism allows for coherence, but it does not ensure that an observer's informational record will exhibit it.

Coherence is essential because confirmation in a relational theory must be perspective-indexed. For an observer to use her outcomes as evidence, she must be able to treat them as belonging to one informational thread. If records are structurally incoherent, they cannot serve as evidence for that observer: each outcome becomes epistemically isolated, unable to contribute to hypothesis-testing or to a continuous updating of beliefs. There is no thread suggestive of continuity, no story being told, and thus no way to judge whether an outcome is expected, anomalous, or supportive of a hypothesis. The evidence that updates observer $O$'s credences must be addressable as one data-set by $O$ across $[t_1,t_n]$. If the record $R$ is structurally incoherent for $O$ (i.e. if $\langle H_R \rangle^O_W = 0$), then there is no single history relative to which $O$ can assign non-trivial likelihoods to the joint record and across-time Bayesian conditioning becomes ill-posed for $O$. In contrast, structural coherence ensures that outcomes hang together within a single diachronic perspective, making them apt for empirical confirmation; $\langle H_R \rangle^O_W \neq 0$ allows a single-perspective likelihood over the whole record. Coherence is therefore a condition for the emergence of the I-observer, fulfilling conditions 2 and 3 for observerhood.

\subsection{Sequential weak values reveal coherence}

The next question is operational: how can we assess whether a record is coherent? We argue that sequential weak values (SWVs) are the required tool to do so.

Weak values were introduced in a 1988 article \cite{aharonov1988result}, and have led to a large number of paradoxes arising from applications of weak measurements \cite{aharonov2013quantum,danan2013asking,vaidman2013past,vaidman2014tracing,aharonov2017case,das2020can,liu2020experimental,aharonov2021dynamical,okamoto2023experimentally}, and controversy regarding the interpretation of weak values \cite{alves2017achieving,ferrie2014result,ferrie2014weak,ferrie2015ferrie,gross2015novelty,knee2014amplification,knee2016weak,reznik2023photons, vaidman2017weak,vaidman2017beyond,matzkin2019weak,korotkov2001continuous,lundeen2012procedure,strubi2013measuring,xu2013phase,zhou2013weak,jayaswal2014observing,magana2014amplification,lyons2015power,zhang2015precision,hallaji2017weak}. This present article does not explicitly comment on this controversy; instead it focuses on what weak values can do for us in the current scenario.

SWVs are particularly well-suited for assessing coherence across time for several reasons.

\subsubsection*{Argument A: SWVs are non-invasive and defined between boundary conditions.}

Weak measurements are designed to minimise disturbance relative to strong projective measurements. In the weak-coupling regime, the interaction between the system and pointer can be made sufficiently small such that the resulting disturbance is negligible to leading order, while still allowing information to be extracted from statistical correlations.

A standard discrete weak measurement is performed by conditioning on the pre- and post-selected states of the system. The standard weak value $A_W$ is defined for a system with operator $A$ and pre- and post-selections, $\ket{\psi}$ and $\ket{\phi}$ respectively \cite{aharonov2008two}:

\begin{equation}
    A^O_W=\frac{\bra{\phi}A\ket{\psi}}{\langle \phi|\psi\rangle}.
    \label{eq:weakvaluetsvf}
\end{equation}

In the context of RQM, both the pre-selected and post-selected states are defined relative to observer $O$, and correspond to states of the observer's informational register associated with the relevant record structure. The same goes for the weak value itself, as indicated in equation~\eqref{eq:weakvaluetsvf}. This observer-relative definition is crucial in a framework where absolute facts are denied, and all physical descriptions are relational.

As a result of their perturbative character in the weak-coupling limit, weak measurements can be performed in sequence, enabling access to temporal structure while inducing only a perturbative disturbance of order $O(g^2)$ per measurement in the weak-coupling regime $g \rightarrow 0$. This is essential for investigating the coherence of records over time, since any disturbance introduced by the measurement would undermine the very coherence we aim to assess.

A sequential weak value (SWV) is similarly found by conditioning on the pre- and post-selected states of the system, but for operators $A_1...A_N$, spanning multiple discrete times. The SWV is defined as follows, here relative to observer $O$ \cite{mitchison2007sequential}:

\begin{equation} \label{SWV eqn}
\langle A_n,...,A_1 \rangle^O _W=\frac{\bra{\phi}A_n...A_1\ket{\psi}}{\langle \phi|\psi\rangle}.
\end{equation}

Crucially, equation~\eqref{SWV eqn} is equivalent to equation~\eqref{eq:projformswv} since $H_R = \Pi_{r_n}(t_n) \dots \Pi_{r_1}(t_1)$, which is a set of operators spanning multiple discrete times associated with outcomes $r_1,\dots,r_n$. Equation~\eqref{eq:projformswv} is thus an SWV, meaning that a non-zero SWV is an operational criterion for structural coherence, as per the coherence criterion defined in Section \ref{subsec:coherentrecords}.

SWVs therefore provide the appropriate tool: they allow us to test whether an observer's records across time yield a coherent history without collapsing the system or erasing the very structure we aim to probe.

\subsubsection*{Argument B: SWVs assess structural coherence \textit{across} time.}

SWVs are required to assess across-time coherence because they evaluate record sequences as unified histories rather than as collections of independent snapshots \cite{terris2025weak}. Formally, an SWV applies to the entire history operator $H_R = \Pi_{r_n}(t_n)\cdots\Pi_{r_1}(t_1$ associated with a record $R=\{r_n(t_n),\ldots,r_1(t_1)\}$, and evaluates the joint compatibility of that record between fixed boundary conditions $(|\psi\rangle_O,|\phi\rangle_O)$. This is not equivalent to assessing each projector in isolation, since the weak values of individual projectors do not determine the SWV of their ordered product. In general,

\begin{equation}
\frac{\langle\phi|BA|\psi\rangle}{\langle\phi|\psi\rangle}
\;\neq\;
\Big(\frac{\langle\phi|B|\psi\rangle}{\langle\phi|\psi\rangle}\Big)
\Big(\frac{\langle\phi|A|\psi\rangle}{\langle\phi|\psi\rangle}\Big),
\end{equation}

hence SWVs are intrinsically non-factorisable. This non-factorisability reflects the fact that SWVs implement a diachronic conjunction: a record is affirmed only if all its elements are jointly compatible across time within a single history.

Because the projectors are taken in the Heisenberg picture, the system dynamics are fully encoded in the history operator. In the Schr\"odinger picture, the SWV of $H_R$ takes the form

\begin{equation}
\langle\phi|\,\Pi_{r_N}U_{N,N-1}\Pi_{r_{N-1}}\cdots U_{2,1}\Pi_{r_1}U_{1,0}\,|\psi\rangle,
\end{equation}

where the unitaries $U_{j+1,j}=U(t_{j+1},t_j)$ implement the actual time evolution between records. Each projector acts as a filter: if the evolved state has no support in the subspace selected at any stage, the entire amplitude vanishes. A non-zero SWV therefore signals that the record sequence can be embedded within a single relational history compatible with the relevant boundary conditions. In this sense, an SWV delivers a single verdict on the coherence of the whole sequence, rather than multiple independent answers to single-time questions. For example, for $R=\{\uparrow(t_1),\downarrow(t_2),\uparrow(t_3)\}$, the SWV

\begin{equation}
\langle H_R \rangle_W^O
=
\frac{\langle \phi| \Pi_{\uparrow}(t_3)\,\Pi_{\downarrow}(t_2)\,\Pi_{\uparrow}(t_1) | \psi \rangle}{\langle \phi|\psi \rangle}
\end{equation}

asks whether this entire record is jointly compatible for observer $O$.

This across-time structure is not an interpretive stipulation; it is already enforced by the interaction dynamics of weak measurements themselves. Consider two weak measurement interactions at distinct times $t_1$ and $t_2$, coupling observables $A$ and $B$ to independent pointer degrees of freedom $p_1$ and $p_2$, with interaction Hamiltonian

\begin{equation}
H_{\mathrm{int}}(t)
=
g_1\,\delta(t-t_1)\,A\otimes p_1
+
g_2\,\delta(t-t_2)\,B\otimes p_2 .
\end{equation}

The resulting unitary evolution is

\begin{equation}
U
=
e^{-ig_2 B(t_2)\otimes p_2}
\,e^{-ig_1 A(t_1)\otimes p_1}.
\end{equation}

Expanding to second order in the coupling constants and retaining the leading terms yields

\begin{equation}
U \approx
1
- i g_1 A(t_1)\otimes p_1
- i g_2 B(t_2)\otimes p_2
- g_1 g_2\, B(t_2)A(t_1)\otimes p_2 p_1 .
\end{equation}

The final term is a cross-term arising from interference between the two weak interactions. It is the only contribution that simultaneously depends on both measurement times and contains the ordered product of observables.

Upon pre- and post-selection of the system in states $\ket{\phi}$ and $\ket{\psi}$, this cross-term generates pointer correlations proportional to

\begin{equation}
\frac{\langle \phi|B(t_2)A(t_1)|\psi\rangle}{\langle \phi|\psi\rangle},
\end{equation}

which is precisely the sequential weak value of $A$ at $t_1$ followed by $B$ at $t_2$ for these pre- and post-selected states. Thus, the same ordered product that appears in the history operator and underwrites diachronic conjunction at the logical level arises directly from the unitary dynamics as an interference effect.

The logical non-factorisability of SWVs and their dynamical origin in cross-terms of the unitary evolution therefore express the same underlying structure. Across-time coherence is neither reducible to, nor recoverable from, single-time weak values: it is an intrinsically multi-time feature that must be assessed at the level of histories. Sequential weak values provide exactly this assessment, supplying an operational criterion for whether a record sequence can be embedded in a single coherent history for an observer.

\subsubsection*{Argument C: SWVs are time-symmetric and structurally atemporal.}

It is important to distinguish the operational procedure used to obtain a sequential weak value from the structural interpretation of the quantity itself. Operationally, sequential weak values are obtained through a temporally ordered sequence of weak interactions followed by post-selection. The procedure therefore unfolds in time and is not itself atemporal.

The quantity obtained from this procedure, however, is naturally interpreted as referring to a structure defined across a temporal interval rather than to a property localised at a single instant. Because the SWV is conditioned on both initial and final boundary conditions, it is sensitive to the global structure connecting those conditions rather than to any one momentary state of affairs. In this sense, SWVs may be regarded as structurally atemporal: they characterise relations extending across time rather than events occurring at a particular discrete point in time.

This bears similarities to time-symmetric and global-constraint approaches to quantum theory, in which explanations are framed in terms of consistency between boundary conditions rather than exclusively in terms of dynamical propagation from past to future \cite{aharonov2008two,price1996time,wharton2018new,adlam2022two,sep-qm-retrocausality}. Such approaches are often described as `all-at-once' descriptions, since the relevant explanatory structure is distributed across an entire spacetime process rather than built up sequentially from local causal influences. While we do not commit ourselves here to any particular retrocausal or global-constraint interpretation of quantum mechanics, SWVs naturally inherit some of these structural features through their dependence on both pre- and post-selection.

The time-symmetric character of SWVs follows directly from their construction. Since both boundary conditions contribute to the value, neither temporal endpoint enjoys a privileged explanatory status. The sequence $A(t_1) \rightarrow B(t_2)$ is not a causal chain, but a logical structure determined by global constraints (the boundary conditions). The lack of a privileged temporal direction means that SWVs are not localised to any discrete moment, nor do they imply that one fact causes the next in a classical sense. They are atemporal traces: mathematical projections of coherent paths through the quantum formalism, conditioned on both pre- and post-selection.

This structural perspective is particularly relevant for the present proposal. Since the I-observer is not localised at a single moment but spread across time, we require a tool by which the I-observer can ensure the coherence of records across the relevant temporal span. This tool should be sensitive to temporal relations without being tied to any discrete moment. Because SWVs evaluate record sequences relative to both initial and final conditions, they provide precisely this kind of tool. They probe the internal structure of informational sequences without collapsing into a single measurement event. In this sense, the structural atemporality of SWVs mirrors the time-delocalised character of the I-observer herself.


\subsubsection*{Argument D: SWVs enable the emergence of the I-observer.}

When a sequence of records satisfies the conditions for coherence, as revealed by a non-zero SWV, the I-observer emerges. The I-observer is not an additional entity added to the theory, but a natural consequence of the coherent structure over time.

This coherence is not something that is imposed from the outside or verified after the fact. It is an internal structural feature of the records themselves. SWVs provide an operational means of assessing this structure. Crucially, the SWV does not test the coherence of a pre-existing sequence. It actually expresses whether such a sequence can be coherently embedded in a narrative structure. Thus, SWVs do not require an I-observer to already exist as they are the very condition under which the I-observer becomes meaningful. The I-observer is an emergent role that depends on the coherence itself. This is what makes the I-observer time-delocalised: it is not localised to any one event, but spans the interval between the boundary conditions, held together by this coherent structure.

Therefore, by grounding the I-observer in the coherence of records (demonstrated by SWVs), we provide a mechanism by which observations can function as evidence and contribute to empirical confirmation.

\subsubsection*{Limitations}

While SWVs offer a compelling structural tool for probing temporal coherence, it is important to clarify the limitations of what they capture. SWVs reveal whether a coherent informational structure exists across time, but they may not, on their own, exhaust all the dimensions involved in the emergence of the I-observer.

First, the SWV formalism depends on well-defined initial and final conditions. This reliance might appear limiting in cases where such boundary conditions are not specified. However, this is not a flaw of the formalism, but a reflection of the relational nature of quantum descriptions in RQM. The I-observer, as constructed here, is not defined dependent on the context. She emerges from the presence of a relational frame, defined by these boundary conditions. So, rather than seeing boundary dependence as a weakness, it can be understood as a structural feature of the relational model.

Second, the SWV approach does not explicitly model the physical stability of records, for instance how memory is retained or protected from decoherence. However, SWVs are not intended as a dynamical model of memory storage. They assess whether a hypothetical informational sequence is coherent under the assumption that such records exist. The point is not that SWVs explain memory stability, but that they provide a condition for when the records could, in principle, support an observer-like structure.

Third, SWVs focus on structural coherence and do not address all dimensions of observation. The framework developed here is concerned with whether records generated at different moments can be organised into a unified informational structure that supports the accumulation and assessment of evidence. Its primary aim is to characterise the conditions under which observations can be related to one another in a way that permits reliable inference.

At the same time, this approach does not seek to explain the subjective character of temporal awareness. Questions about how a subject experiences continuity across time, or what underlies the sense of being the same individual from one moment to the next, fall outside the scope of the present analysis. The proposal is therefore intentionally limited: it identifies a structural requirement for observation as an evidential practice without claiming to provide a comprehensive theory of consciousness, selfhood, or lived temporal experience.

This limitation does not diminish the relevance of the framework for the problem under consideration. Before observations can contribute to the evaluation of hypotheses, they must be preserved, related, and compared within a sufficiently stable informational organisation. SWVs offer a formal means of assessing this prerequisite. Although further theoretical work may be required to account for the experiential dimension of observation, the ability to maintain coherent records remains a fundamental condition for their use in generating and assessing evidence.

SWVs therefore identify a minimal structural condition for the evidential role of observation. Their value lies in evaluating whether records can be integrated into a coherent informational framework capable of supporting comparison, inference, and confirmation across time. Rather than undermining the account, the limitations discussed above help to clarify its scope: the present proposal is not intended as a theory of consciousness or subjective persistence, but as an account of the informational architecture required for observations to function as evidence. In this sense, SWVs delineate the structural conditions within which the emergence of the I-observer can occur.

\section{Agreement and Confirmation in the Wigner's Friend Paradox}\label{Sec:two-step}

The preceding sections developed a structural criterion for observerhood in RQM, grounded in the coherence of records across time and operationally assessed by SWVs. We have argued that an observer exists, in the full epistemic sense required for scientific practice, only when her records can be embedded into a single diachronic history relative to appropriate boundary conditions. A non-vanishing SWV functions as a coherence check: it certifies that the informational role of the observer is well-defined across the relevant temporal interval.

In this section, we demonstrate how this coherence criterion operates in a setting where relational descriptions are known to come into tension: the Wigner's friend scenario \cite{wigner1995remarks}. Wigner's friend–type experiments make evident the problem of intersubjective agreement in RQM. If coherence across time cannot be secured here, it is unclear how RQM could support agreement and hence confirmation at all.

The aim of this section is threefold. First, we review the Wigner's friend paradox and the conditions under which observers cannot agree on the outcome of a measurement. We discuss existing approaches to this problem and the difficulties they face in recovering empirical confirmation within a fully relational framework. Second, we show how the SWV coherence criterion developed in the previous section establishes the structural coherence of the friend's records, allowing them to support a stable evidential perspective and thereby giving rise to the I-observer. We then apply this coherence criterion within the RQM+CPL framework \cite{adlam2022information}, demonstrating how a coherent record can subsequently be accessed by Wigner, enabling intersubjective agreement without reintroducing observer-independent facts. These results suggest a route to empirical confirmation within RQM while preserving its fundamentally relational character. Third, we examine the Frauchiger--Renner paradox as a limiting case. This paradox raises a more demanding challenge involving multiple nested observers and mutually incompatible inferences. Examining this case helps clarify the scope of the coherence criterion and its role in relation to broader questions of intersubjective agreement and empirical confirmation. 

\subsection{Wigner and his friend cannot agree with each other}

\begin{figure}
    \centering
    \includegraphics[scale=0.1]{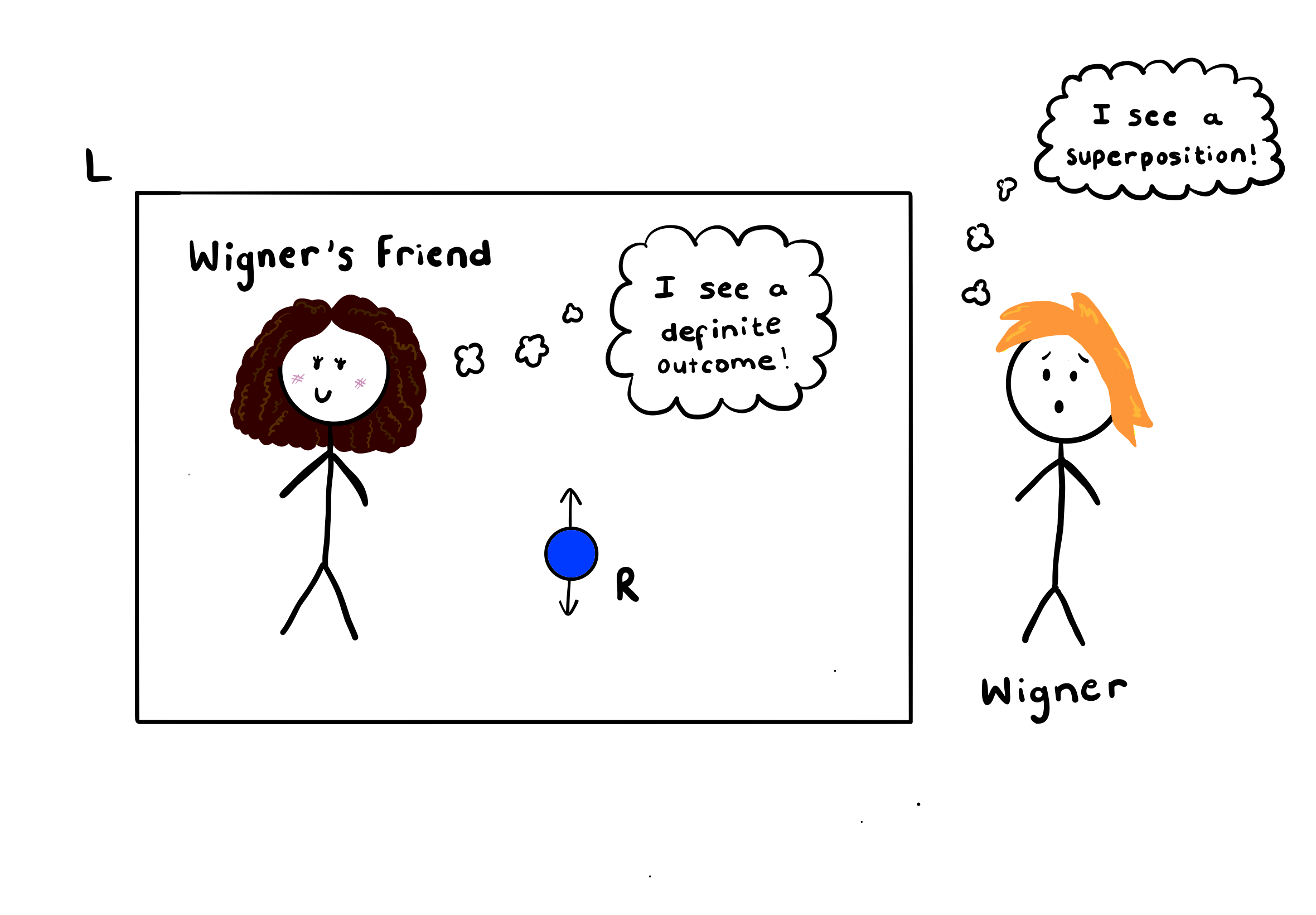}
    \caption{The Wigner's friend paradox: A friend inside of a closed laboratory $L$ measures a qubit and obtains a definite outcome. Wigner is outside $L$ and sees the entire laboratory in a superposition of states. The perspectives of Wigner and his friend appear to contradict each other}
    \label{Fig1}
\end{figure}

The Wigner's friend paradox \cite{wigner1995remarks,deutsch1985quantum} imagines a closed laboratory, inside of which is a qubit and an observer (the friend), as depicted in Figure \ref{Fig1}. The qubit is prepared in the superposition state $\frac{1}{\sqrt{2}}(\ket{\uparrow}+\ket{\downarrow})_R$. Unitary quantum mechanics is assumed.

Inside the laboratory, the friend performs a measurement on the qubit, collapsing the superposition such that from her perspective, the system is now in a definite state, either $\ket{\uparrow}_R$ or $\ket{\downarrow}_R$. Meanwhile, Wigner is waiting outside the closed laboratory. He cannot communicate with his friend at this stage, so he has no information regarding the friend's measurement outcome. From Wigner's perspective, the qubit is in the superposition state in which it was initially prepared. According to Wigner, the entire laboratory is in a superposition state since the friend's ready state $\ket{0}_F$ is entangled with the preparation state of the qubit: $\ket{\Psi}_W=\ket{0}_F\otimes\frac{1}{\sqrt{2}}(\ket{\uparrow}+\ket{\downarrow})_R$.

The original Wigner's friend scenario is primarily a conceptual thought experiment. It highlights the tension between the friend's definite outcome and Wigner's unitary description of the laboratory, thereby raising questions about the status of facts and agreement between observers. It raises the paradox: if quantum mechanics is universally valid, why do Wigner and his friend hold different perspectives on the state of the same system at the same time? 

More recent developments have sharpened these concerns by embedding Wigner's friend–type situations within formal no-go frameworks. One important example is the Local Friendliness theorem \cite{bong2020strong}, which investigates the compatibility of observer-independent events with additional assumptions concerning locality and freedom of measurement settings.

While Local Friendliness is motivated by Wigner's friend scenarios, it provides a more precise framework for identifying which assumptions must be abandoned when observer-dependent descriptions are taken seriously. It is therefore useful here, since it makes explicit the tensions surrounding observer-independent facts that motivate relational approaches such as RQM.

Local Friendliness consists of the conjunction of the following three assumptions:

\hfill

\begin{adjustwidth}{0.5cm}{0.5cm}

\textbf{ABSOLUTENESS OF OBSERVED EVENTS (AOE):} An observed event is a real, single event, and not relative to anything or anyone.

\hfill

\textbf{NO-SUPERDETERMINISM:} Any set of events on a space-like hypersurface is uncorrelated with any set of freely chosen actions subsequent to that space-like hypersurface.

\hfill

\textbf{LOCALITY:} The probability of an observable event is unchanged by conditioning on a space-like-separated free choice [...].

\end{adjustwidth}

\hfill

The Wigner's friend paradox suggests a tension between these assumptions, whereby we cannot maintain all three at once. RQM provides a solution by rejecting AOE, allowing the different perspectives to be equally valid within their own relational contexts. The friend can then assign a definite state to the qubit due to her interaction with it. Meanwhile, as Wigner has not yet interacted with the qubit nor with his friend, he can unproblematically conclude that the state of the qubit-friend system remains in a superposition. There is no contradiction between the different perspectives of the agents, as the descriptions are relative to each respective agent and are each valid within their own contexts.

However, this solution presents the following problem: if there is no absolute perspective, and no way for Wigner and the friend to reconcile their observations, how can there be any agreement between observers? Without intersubjective agreement, RQM risks not being empirically confirmable.

The problem worsens if at some later time Wigner is allowed to open the laboratory and communicate with his friend, asking `What did you see?' The friend may answer `I saw spin-up', communicating her outcome to Wigner. But according to RQM, this generates a fact relative to Wigner: `My friend saw spin-up'. Crucially, although both observers may assign the same outcome value (`spin-up'), the corresponding relational facts are not identical. Wigner's fact concerns the friend's recorded outcome relative to Wigner, whereas the friend's fact concerns her own measurement outcome relative to herself.

In other words, Adlam argues that the way in which RQM treats communication as the generation of relative facts leads to a fundamental disagreement at the heart of the theory \cite{adlam2022does}. To see this, let us denote $M^S_W$ as the outcome of the system measurement as inferred by Wigner, and $M^F_W$ as the friend's measurement outcome as reconstructed by Wigner through his interaction with the friend. Suppose also that it is really the outcome measured and recorded by the friend within her own reference frame. RQM allows for a scenario in which:

\begin{equation}
    M^S_W = M^F_W, \text{but } M^F \neq M^S_W.
\end{equation}

This entails:

\begin{equation}
    M^F \neq M^F_W.
\end{equation}

That is, from Wigner's perspective, the friend's declared measurement outcome (as inferred via an interaction between Wigner and the friend) may agree with the system's behaviour as Wigner has observed it, but it may still fail to coincide with what the friend herself takes to have observed.

While there have been some attempts to recover agreement between agents in the RQM framework, these attempts face significant difficulties as they counter the rejection of absolutism central to the interpretation. One response developed within relational and perspectival approaches argues that apparent disagreement between observers arises only when descriptions are illegitimately compared from an absolute standpoint. On this view, whether two observers agree is itself a relational question: Alice's account of Bob's facts relative to Alice may align with Bob's account of his own facts, even if there is no observer-independent fact of the matter about their agreement \cite{brown2009relational,calosi2024relational}. Such approaches preserve the relational core of RQM by treating perspectival alignment itself as observer-relative. However, the present article is motivated by a different concern: even if perspectival consistency is secured, scientific observation and empirical confirmation still appear to require that records be retained, related, and treated as belonging to a coherent temporally extended evidential perspective.

A second strategy appeals to \emph{shared facts}, arguing that when measurement outcomes become stable through decoherence, they can be treated as effectively irreversible records that different observers will later agree upon \cite{di2021stable,pienaar2021quintet}. However, this proposal either restricts agreement to the trivial case in which observers directly measure the same pointer variable, or else violates RQM’s requirement that different perspectives cannot be directly compared without mediation by a third interaction.

A third proposal referred to in the literature as \emph{internally consistent descriptions} avoids this conflict by restricting consistency claims to what a single observer can jointly describe, requiring only that an observer who measures both a system and another agent in the same context finds the results to agree \cite{di2022relational}. Yet this guarantees agreement only in cases where perspectives are already aligned, and provides no way to reconcile genuinely divergent descriptions \cite{adlam2022does}.

Finally, \emph{cross-perspective links} (CPL) posits that an agent's measurement outcome is encoded in physical memory variables that persist through time and can later be accessed by other agents, thereby allowing agreement via the direct readout of those records \cite{adlam2022information}. In this way, CPL provides a concrete mechanism for relating descriptions across observers and represents one of the most substantial attempts to recover intersubjective agreement within RQM. Importantly, CPL explicitly restricts itself to cases in which the relevant records remain sufficiently stable and undisturbed for cross-perspective accessibility to obtain. The complementary model of the observer developed in Section \ref{Sec:rethinking}, combined with the coherence-check outlined in Section \ref{Sec:coherent}, supplements this framework by developing a structural criterion for when temporally extended records may be treated as informationally coherent within a relational setting. Sequential weak values are not intended to provide a dynamical guarantee of physical memory preservation, but rather to assess whether a record sequence can be embedded into a coherent informational perspective across time. The following subsection demonstrates this application of our proposal to RQM$+$CPL, illustrating how coherent records may support intersubjective agreement between Wigner and his friend while preserving the relational structure of the theory.

\subsection{Two-step confirmation model}

Consider the original Wigner's friend scenario. Let $\mathcal{H}_S$ denote the Hilbert space of the measured qubit and $\mathcal{H}_M$ the Hilbert space associated with the friend's memory degrees of freedom. The qubit is initially prepared in the superposition
\[
\ket{\psi}_S=\frac{1}{\sqrt{2}}(\ket{\uparrow}_z+\ket{\downarrow}_z),
\]
and the friend's measurement interaction correlates the qubit state with corresponding memory states in $\mathcal{H}_M$. The relevant coherence criterion therefore concerns the informational structure encoded in the friend's memory record rather than the isolated state of the qubit itself. For simplicity, the notation suppresses the explicit tensor-product structure of the composite system. The pre- and post-selected states should therefore be understood as encoding boundary conditions on the friend's informational record, rather than as states of the isolated qubit alone.

Suppose that, relative to the friend's perspective, the relevant informational boundary conditions correspond to a coherent record associated with the outcome $\ket{\uparrow}_z$. The corresponding sequential weak value is then:
\begin{equation}
\langle\sigma_z(t_3,t_2,t_1)\rangle_F
=
\frac{\bra{\phi}\sigma_z(t_3)\sigma_z(t_2)\sigma_z(t_1)\ket{\psi}}{\langle \phi|\psi\rangle}
=1.
\end{equation}

This quantity evaluates whether the sequence of records associated with the friend's informational register across the interval $[t_1,t_3]$ can be embedded in a single coherent history compatible with her boundary conditions. A vanishing value would indicate that the recorded outcomes cannot be jointly realised within one relational narrative. The non-zero result, by contrast, affirms that the friend's records form a coherent informational structure.

The friend's records may therefore be treated as forming a single coherent informational perspective across the relevant interval. What is confirmed is not the metaphysical permanence of the memory state, but the logical and structural integrity of the narrative it constitutes over the relevant interval.

This first step therefore establishes that the friend's records form a coherent informational structure capable of supporting a stable evidential perspective across time. What is secured is not a new piece of knowledge acquired by the friend, but the structural condition under which those records can function as evidence within her perspective. However, this alone does not secure agreement with Wigner. The coherence established here remains perspective-indexed. Nothing yet guarantees that Wigner can access, interpret, or verify the friend's record.

A second step is therefore required. Working within RQM+CPL, Wigner later opens the laboratory and measures the composite system, including the friend's memory register. From Wigner's perspective, the laboratory has evolved unitarily. The CPL postulate ensures that if the friend's outcome is encoded in stable physical degrees of freedom, then Wigner's measurement of those degrees of freedom yields a result matching the friend's recorded outcome. Symbolically,
\[
M^F_W = M_F.
\]

This does not produce an observer-independent fact about the qubit. Wigner's result remains relational. Rather, it establishes agreement between perspectives: Wigner confirms that the outcome he reads from the friend's memory corresponds to the outcome genuinely recorded by her.

The two-step procedure can therefore be summarised as follows. First, the SWV coherence criterion is satisfied, indicating that the friend's records form a coherent informational structure capable of supporting a single observer-perspective across the relevant temporal interval. Second, CPL enables Wigner to access that record and establish agreement across observers.

The two steps address different aspects of the confirmation problem. CPL specifies how information encoded in one observer's memory can later be accessed by another observer, thereby enabling agreement between perspectives. However, CPL presupposes that the relevant record remains a coherent evidential structure across time. The role of the SWV coherence criterion is to assess precisely this prior requirement. In this sense, the present proposal supplements rather than replaces CPL. Whereas CPL concerns the accessibility of records across observers, the coherence criterion concerns whether those records can function as a coherent evidential perspective in the first place.

This model therefore preserves the relational core of RQM while demonstrating how empirical confirmation can be recovered in Wigner's friend scenarios. Agreement remains perspective-indexed, yet stretches across observers. Within RQM+CPL, coherent records and cross-perspectival accessibility together ground meaningful intersubjective agreement without appealing to observer-independent facts. The present proposal therefore offers a route to recovering empirical confirmation within RQM, given CPL and the proposed coherence criterion.

This example is intended as a schematic illustration of the coherence criterion rather than as a fully realistic model of the Wigner's friend experiment. The purpose of the calculation is to show how a coherent record sequence is represented within the formalism. More complex behaviour arises when different boundary conditions are considered, as illustrated in the Frauchiger-Renner case discussed in Section \ref{subsec:FR}.


\subsection{Frauchiger-Renner as a limit case for confirmation}\label{subsec:FR}

A contrasting case arises in extended Wigner's friend scenarios such as the Frauchiger-Renner paradox \cite{frauchiger2018quantum}. The scenario involves two `friends', $F$ and $F'$, each located inside an isolated laboratory, together with two external observers, $W$ and $W'$, who later perform measurements on those laboratories. The friends first perform measurements and record definite outcomes relative to themselves. The external observers subsequently describe the laboratories as quantum systems evolving unitarily and perform measurements on the laboratories as wholes. By combining the reasoning of all four agents, it is shown that each observer can derive conclusions about the outcomes of the others that appear individually valid, yet cannot all be jointly maintained. The paradox therefore exposes a tension between different observer-perspectives and has been widely interpreted as a challenge for interpretations that seek to maintain both universal unitary evolution and consistent reasoning across agents.

This makes Frauchiger-Renner a useful test-case for the present proposal, since it allows local informational coherence to persist while broader cross-perspectival coherence breaks down.

Consider observer $F$ in laboratory $L$. Relative to $F$, the qubit received from $F'$ is initially described by the pre-selected state
\[
\ket{\psi}_F
=
\frac{1}{\sqrt{2}}
(
\ket{\uparrow_S}
+
\ket{\downarrow_S}
)_F.
\]
Suppose now that the post-selection corresponds to the state
\[
\ket{\phi}_F=\ket{\downarrow_S}_F.
\]
The associated sequential weak value is then
\begin{equation}
    \begin{split}
        \langle\sigma_z(t_3,t_2,t_1)\rangle_F
        &=
        \frac{
        \bra{\downarrow_S}
        \sigma_z(t_3)\sigma_z(t_2)\sigma_z(t_1)
        \ket{
        \frac{1}{\sqrt{2}}
        (\ket{\uparrow_S}+\ket{\downarrow_S})
        }_F
        }{
        \bra{\downarrow_S}
        \frac{1}{\sqrt{2}}
        (\ket{\uparrow_S}+\ket{\downarrow_S})
        \rangle_F
        }
        \\
        &=-1.
    \end{split}
\end{equation}

The significance of this result is subtle. The sequential weak value does not vanish, and so the records relative to $F$ still exhibit local informational coherence within that observer-perspective. In this sense, the record sequence remains structurally coherent for $F$, allowing her outcomes to function as part of a single diachronic evidential perspective. However, the anomalous value reflects the contextual nature of that coherence. While the records remain coherent relative to $F$, they resist incorporation into a single globally coherent informational structure spanning all observer-perspectives within the Frauchiger-Renner scenario. The resulting tension is therefore not one of simple logical contradiction or total coherence failure, but of incompatible contextual narratives that cannot all be jointly embedded within one unified relational description.

From the perspective developed in the present article, this illustrates that temporal coherence operates at multiple levels. A sequential weak value of zero signals the breakdown of coherence within a single observer-perspective, whereas the Frauchiger-Renner scenario instead reveals a limitation on the coherent integration of multiple perspectives despite the persistence of local coherence. Individual observers may therefore retain internally coherent evidential narratives while broader intersubjective coherence nevertheless breaks down. In such cases, empirical confirmation becomes fragmented across perspectives: records remain locally meaningful, but cannot straightforwardly function as components of a single mutually coherent evidential framework.

\section{Conclusion}\label{Sec:conc}

This article has addressed a central difficulty for RQM: how empirical confirmation can be sustained within a framework that defines any physical system as an observer. Without further criteria, such a minimal definition does not by itself provide a way to determine when an observer's records constitute a stable evidential perspective across time. If outcomes cannot be treated as belonging to a single, coherent, time-extended standpoint, they cannot function as evidence, and observerhood itself becomes epistemically fragile. A related challenge concerns intersubjective agreement: in the absence of observer-independent facts, it is unclear how distinct perspectives can be brought into evidential alignment. We have argued that both problems can be resolved using our two-step model.

By reconceiving the observer as a complementary two-part physical and informational structure and introducing a coherence criterion grounded in sequential weak values, we identify the conditions under which records form a coherent structure. This coherence allows outcomes to function as evidence for that observer. This coherence criterion does not merely precede cross-perspective links; it supplies the missing justification for treating records as stable carriers of information across time. Once this stability is established, CPL can operate non-circularly: coherent records can be accessed and verified by another observer without presupposing the very persistence they are meant to confirm. Empirical confirmation is thus recovered not as global agreement, but as a relational achievement grounded in structurally coherent perspectives.

The resulting framework remains relational in spirit, but it is not identical to standard formulations of RQM. Rovelli's original account deliberately avoids introducing a privileged class of observers, treating observation as a generic physical interaction \cite{rovelli1996relational}. By contrast, the present proposal argues that such a minimal notion of observerhood is insufficient for the purposes of empirical confirmation. If records are to function as evidence, observers must satisfy an additional coherence criterion across time. In this sense, the present account should be understood not merely as an application of RQM, but as a supplement to it: the relational character of facts is preserved, while the requirements placed on observerhood are strengthened. This conclusion also resonates with recent criticisms of unrestricted observerhood in relational approaches, particularly Brukner's argument that qubits should not automatically qualify as observers \cite{brukner2021qubitsobserversnogo}.

This model also resonates with broader philosophical perspectives. In particular, it shares important themes with structural realist approaches \cite{ladyman2007every,french2014structure}, insofar as observerhood is characterised in terms of relational and informational structure rather than intrinsic properties. The I-observer is not introduced as a fundamentally distinct entity, but emerges from patterns of coherence that hold records together across time. While the present argument does not establish the stronger metaphysical claims associated with ontic structural realism, it is nevertheless compatible with the idea that relational structure plays a constitutive role in the emergence of observers.

Several avenues for future research arise from our account. First, our account invites further investigation into the relation between observer coherence and indefinite causal structure. In settings where causal order is not well-defined, temporally extended coherence may become even more important. Because if there is no global time, coherence may be the only structure capable of defining an observer.

Second, this account raises broader questions regarding the philosophy of science. If confirmation is always internal and relational, then we may need to rethink the structure of scientific confirmation itself. Rather than thinking of science as converging on a single, global truth, we may instead think of it as a network of internally coherent perspectives. Such a view bears some affinity to discussions of situated knowledge and social objectivity in feminist philosophy of science, particularly Longino's emphasis on the critical interaction between diverse perspectives as a source of objectivity \cite{longino1990science,sep-scientific-knowledge-social}. While the present account is developed within the context of quantum theory rather than social epistemology, both approaches challenge the idea that objectivity requires access to a single perspective-independent standpoint.

Ultimately, the central lesson of this work is that empirical confirmation depends on more than the existence of observer-relative facts. It requires observers whose records form coherent structures across time. By providing a criterion for that coherence, sequential weak values help bridge the gap between relationality and confirmation, allowing RQM to retain its relational character without sacrificing its empirical foundations.

\backmatter

\bmhead{Acknowledgements}
The author extends gratitude to Alexei Grinbaum for supervision and for numerous
discussions on the notion of the observer.

\bmhead{Author Contributions}
The author confirms sole responsibility for the following: conception of analysis,
design of all figures, preparation of the manuscript.

\bmhead{Funding}
This research was funded, partially, by l’Agence Nationale de la Recherche (ANR), project
ANR-22-CE47-0012.

\section*{Declarations}

\bmhead{Conflict of interest}
The authors declare no competing interests.

\bibliography{sn-bibliography}

\end{document}